\documentstyle[float,twocolumn,prl,aps,epsfig]{revtex} 

\newcommand{\nc}{\newcommand}
\nc{\bb}{\bbox{b}}
\nc{\bs}{\bbox{s}}
\nc{\pt}{p_{\rm T}}
\nc{\mt}{m_{\rm T}}
\nc{\pL}{p_{\rm L}}
\nc{\ET}{E_{\rm T}}
\nc{\Nch}{N_{\rm ch}}
\nc{\Nc}{N_{\rm coll}}
\nc{\Np}{N_{\rm part}}
\nc{\se}{\section}
\nc{\suse}{\subsection}
\nc{\beq}[1]{\begin{equation}\label{#1}}
\nc{\eeq}{\end{equation}}
\nc{\bea}[1]{\begin{eqnarray}\label{#1}}
\nc{\eea}{\end{eqnarray}}
\nc{\bce}{\begin{center}}
\nc{\ece}{\end{center}}
\nc{\bit}{\begin{itemize}}
\nc{\eit}{\end{itemize}}
\nc{\bmp}{\begin{minipage}}
\nc{\emp}{\end{minipage}}

\newcommand{\lapp}{\raisebox{-.5ex}{$\stackrel{<}{\sim}$}}

\nc{\la}{\langle}       
\nc{\lla}{\left \langle}
\nc{\ra}{\rangle}       
\nc{\rra}{\right \rangle}
\newcommand{\lda}{\langle\!\langle}       

\newcommand{\rda}{\rangle\!\rangle}

\newcommand{\half}{{\textstyle{1\over 2}}}

\begin{document}

\title{Centrality Dependence of Multiplicity, Transverse Energy, and 
Elliptic Flow from Hydrodynamics}

\preprint{JYFL-4/2001//LBNL-47635//hep-ph/0103234}

\author{P.F.~Kolb$^{a,b}$, U.~Heinz$^a$, P.~Huovinen$^c$, K.J.~Eskola$^{d,e}$, 
        and K.~Tuominen$^d$}
\address{$^a$Department of Physics, The Ohio State University, 174 West
         18th Avenue, Columbus, OH 43210, USA\\
         $^b$Institut f\"ur Theoretische Physik, Universit\"at
         Regensburg, D-93040 Regensburg, Germany\\
         $^c$Lawrence Berkeley National Laboratory, Berkeley CA 94720, USA\\
         $^d$Department of Physics, University of Jyv\"askyl\"a, P.O.~Box 35, 
         FIN-40351 Jyv\"askyl\"a, Finland\\
	 $^e$Helsinki Institute of Physics, P.O. Box 64, FIN-00014 University 
	 of Helsinki, Finland}
\date{Submitted to Nucl. Phys. A on March 23, 2001}

\maketitle


\begin{abstract}
The centrality dependence of the charged multiplicity, transverse energy, 
and elliptic flow coefficient is studied in a hydrodynamic model, using a 
variety of different initializations which model the initial energy or 
entropy production process as a hard or soft process, respectively. While 
the charged multiplicity depends strongly on the chosen initialization, the 
$\pt$-integrated elliptic flow for charged particles as a function 
of charged particle multiplicity and the $\pt$-differential elliptic 
flow for charged particles in minimum bias events turn out to be almost 
independent of the initial energy density profile. 
\end{abstract}

\smallskip



\medskip

\vspace*{-0.3cm}
	
\section{Introduction}
\label{sec1}

\vspace*{-0.15cm}
 
Elliptic flow \cite{O92} is a collective flow pattern which develops
in non-central relativistic heavy-ion collisions as a result of the
spatial deformation of the initial transverse overlap area. It requires
rescattering among the produced particles as a mechanism to map the 
initial spatial deformation of the reaction zone onto the finally
observed hadron momentum distributions. It is quantified by the second
harmonic coefficient $v_2$ of an azimuthal Fourier decomposition of
the measured spectrum $dN/(dy\, \pt d\pt d\phi)$ \cite{VZ96}. Its 
magnitude $v_2$ and its shape $v_2(\pt)$ as a function of the hadron 
transverse momentum are sensitive to the scattering rate among the 
produced secondaries, especially during the dense early stage of the 
expansion \cite{S97,ZGK99}. The largest elliptic flow signal, especially 
at high $\pt$, arises in hydrodynamic simulations \cite{O92,KSH99,KSH00} 
which assume local thermal equilibrium at every space-time point, i.e. 
essentially instantaneous thermalization or infinite scattering rate. 
Surprisingly, such hydrodynamic simulations \cite{KSH99} are in very 
good agreement with first results from $\sqrt{s}=130\,A$\,GeV 
Au+Au collisions at the Relativistic Heavy Ion Collider (RHIC) 
\cite{STAR00,KH3,HKHRV,Talks,QM01}, up to transverse momenta of 
$1.5 - 2$\,GeV/$c$.

It was recently suggested \cite{KH3,HKHRV,TLS00} that a combined ana\-lysis 
of the full set of hadronic single particle spectra and their elliptic 
flow as a function of collision centrality should allow to outline the
domain of applicability of the hydrodynamic approach. Inside this domain
such an analysis would constrain the initial baryon and energy density
and the final freeze-out temperature sufficiently well to become sensitive 
to details of the equation of state (EOS) of the fireball matter 
\cite{HKHRV}. In this context $v_2$ provides access to the EOS during 
the early expansion stage \cite{S97,ZGK99,KSH99} which, at RHIC energies, is 
presumably in the quark-gluon plasma (QGP) phase. Direct verification of 
the phase transition between QGP at high energy density and a hadron 
resonance gas at lower energy density should then become possible by 
accurately measuring the excitation function of radial and elliptic 
flow \cite{KSH99}.

In the hydrodynamic limit the EOS affects the elliptic flow signal 
through the velocity of sound, $c_s=\sqrt{dP/de}$ \cite{O92}. However, the
sensitivity of $v_2$ on $c_s$ is not very strong, and even a first
order phase transition, where $c_s$ vanishes in the mixed phase, 
affects $v_2$ only on the 20$\%$ level \cite{KSH99}. This makes accurate 
measurements and systematic theoretical checks indispensable. One possible 
source of ambiguity, which has not been systematically investigated in 
previous studies \cite{O92,KSH99,KSH00,KH3,HKHRV,TLS00}, is the 
sensitivity of the radial and elliptic flow pattern on the shape of
the initial transverse density profile. The latter depends on the
scaling of secondary particle production with the number of colliding
nucleons which itself is controlled by the collision centrality. It was
recently found \cite{GVW01} that the momentum-space anisotropy at high 
$\pt$, where it is not of hydrodynamic origin but due to quark energy 
loss \cite{W00}, is quite sensitive to the initial density profile. A 
good hydrodynamic baseline for the dependence of $v_2(\pt)$ on the initial
profiles may thus help to clarify the physics controlling the breakdown 
of the hydrodynamic model and the transition to the hard physics domain 
at high $\pt$.

In the present work, we investigate five options which are expected to 
span the realistic range of possibilities. In the first four, either the 
initial energy or the initial entropy density are assumed to scale with 
either the number of wounded nucleons (``soft'' or ``non-perturbative'' 
particle production) or the number of binary nucleon-nucleon collisions 
(``hard'' or ``perturbative'' particle production). In the fifth model 
perturbative particle production is modified by implementing gluon 
shadowing \cite{EKS98} in the initial state and by limiting the growth 
of the production cross section by gluon saturation in the final state 
(``saturation model'' \cite{EKRT00,EKT01}). This brings in some 
non-perturbative elements as well. In the first four parametrizations 
we normalize the initial energy density profile such that for central 
collisions we reproduce within errors the total charged multiplicity 
density at midrapidity,
$d\Nch/d\eta\vert_{\vert\eta\vert<1}{\,=\,}555\pm10\%$, as measured by
PHOBOS for Au+Au at $\sqrt{s}{\,=\,}130\,A$\,GeV \cite{PHOBOS}. On the
other hand, in the saturation model \cite{EKRT00,EKT01} the initial
energy density is fixed by assuming that the transverse energy of
produced minijets is entirely converted to thermalized energy
density. The different scaling laws implied by the models then
translate into different centrality dependencies of $d\Nch/d\eta$
which can be tested against the new data presented recently
\cite{PHENIX,Roland}.

The hydrodynamic model describes an adiabatic evolution from one local 
equilibrium state to another. Our version of the model assumes longitudinal
boost invariance which implies conservation of the entropy rapidity 
density $dS/dy$. Using the relation between entropy and particle 
multiplicity in a thermalized system, the measured centrality dependence 
of the final multiplicity density $dN/dy$ can thus be mapped onto the 
centrality dependence of the initial parton multiplicity density at the 
point of thermalization. This allows to constrain models for the initial
production of secondary particles, under the assumption of subsequent 
adiabatic evolution.     

We show that different models for the initial energy and entropy 
production lead to different radial and elliptic flow patterns. At 
fixed impact parameter, these result in different $\pt$-dependences 
of the spectra and of the differential elliptic flow $v_2(\pt)$. However, 
the $\pt$-integrated elliptic flow $v_2$ as a function of the final 
charged multiplicity density $d\Nch/dy$ and the differential elliptic 
flow $v_2(\pt)$ for minimum bias events show surprisingly little 
sensitivity to the model used for initializing the hydrodynamic evolution. 
These two observables have been used to argue for the applicability of 
the hydrodynamic model at RHIC energies \cite{STAR00,KH3}, and they were 
shown to be sensitive to the EOS of the expanding matter \cite{KH3,HKHRV}. 
Their insensitivity to the shape of the initial transverse energy density 
profile may prove to be crucial for the process of extracting reliable 
information on the EOS from a detailed analysis of the measured collective 
flow patterns.

\vspace*{-0.2cm}
 
\section{Models for the initial transverse energy density profile}
\label{sec2}
\subsection{Soft particle production}
\label{sec2a}

\vspace*{-0.15cm}

In Pb+Pb collisions at the SPS, the rapidity densities at midrapidity of 
both the total produced transverse energy,\footnote{We use 
$\ET = \sum_i E_i\, {p_{{\rm T},i}\over |\bbox{p}_i|}$ where the sum is 
over all particles.} $d\ET/dy$, and of the charged multiplicity, $d\Nch/dy$, 
scale approximately linearly with the number of participating nucleons, 
$\Np$ \cite{WA98}, except for very peripheral collisions with 
$\Np\lapp 100$. Similar earlier observations in smaller collision systems 
and/or at lower energies have led to the notion that particle and 
transverse energy production can be described by the ``Wounded Nucleon 
Model'' \cite{WNM} in which each participating nucleon contributes to 
the multiplicity and transverse energy only once in its first collision 
and not every time it suffers further collisions with other projectile 
or target nucleons. The validity of this model requires destructive 
interference effects between subsequent nucleon-nucleon collisions 
which are thought to be characteristic of non-perturbative or ``soft'' 
particle production processes. For two nuclei colliding at impact 
parameter $\bb{\,=\,}b\,\bbox{e}_x$, the density of wounded nucleons 
in the transverse plane (parametrized by $\bs{\,=\,}(x,y)$) can be 
calculated from the simple geometric formula (Glauber ansatz, see 
\cite{BO90,KLNS} and references therein):
 \bea{init}
   n_{\rm WN}(\bs;\bb) &=&
   T_A\bigl(\bs{+}\half\bb\bigr)
      \Bigl[1-\Bigl(1-{\sigma T_B\bigl(\bs{-}\half\bb\bigr)
                       \over B}\Bigr)^B \Bigr] 
 \nonumber\\
   &+& T_B\bigl(\bs{-}\half\bb\bigr)
     \Bigl[1-\Bigl(1-{\sigma T_A\bigl(\bs{+}\half\bb)
                      \over A}\Bigr)^A \Bigr]\,.
 \eea

\vspace*{-0.5cm}
\begin{figure}
  \begin{center}
    \epsfxsize 7cm \epsfysize 5.3cm \epsfbox{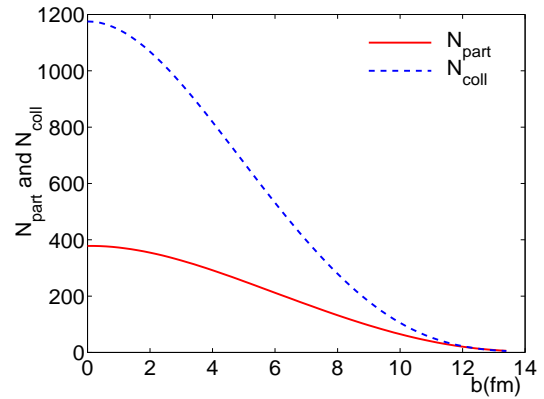}
  \end{center}
\caption{Number of participating (``wounded'') nucleons and of binary 
nucleon-nucleon collisions as functions of impact parameter. This and
all following figures refer to Au+Au collisions at $\sqrt{s}=130\,A$\,GeV. 
}
\label{F1}
\end{figure}

$T_A$ is the nuclear thickness function of nucleus $A$,
 \beq{TA}
   T_A(\bs)=\int_{-\infty}^{+\infty} dz\, \rho_A(\bs,z) \,,
 \eeq
with the density $\rho_A$ given by a Woods-Saxon profile,
 \beq{rhoA}
   \rho_A(\bbox{r}) = \frac{\rho_0}{1+\exp[(r-R_0)/\xi]}\,,
 \eeq
and similarly for nucleus $B$. For the Woods-Saxon profile we take 
standard parameter values, $R_0{\,=\,}1.12\,A^{1/3}-0.86\,A^{-1/3}$\,fm 
for the radius and $\xi{\,=\,}0.54$\,fm for the surface diffuseness.
The integral over (\ref{init}) gives the number of ``wounded'' or 
``participating'' nucleons, $\Np\equiv N_{\rm WN}$, as a function of 
impact parameter (see Fig.~\ref{F1}). For the total nucleon-nucleon cross 
section at $\sqrt{s}{\,=\,}130\,A$\,GeV we take $\sigma{\,=\,}40$\,mb. 


\subsubsection{Model \ {\rm sWN}}
\label{sec2a1}

Hydrodynamics with boost-invariant longitudinal ex\-pansion conserves the 
entropy per unit rapidity $dS/dy$. At fixed freeze-out temperature and 
chemical potential, $dS/dy$ is related one-to-one to the measured charged 
multiplicity density, $d\Nch/dy$. If the latter scales linearly with the 
number of wounded nucleons,
$N_{\rm WN}(b){\,=\,}\int d^2s\,n_{\rm WN}(\bs;\bb)$, barring a strong 
$b$-de\-pen\-dence of the freeze-out conditions it is natural to assume 
that the initial entropy density in the transverse plane is proportional 
to that of the wounded nucleons \cite{O92,TS99}:
 \bea{sw}
   s(\bs;\tau_0;\bb) = K_s(\tau_0) \, n_{\rm WN}(\bs;\bb)\,.
 \eea
At SPS and RHIC energies the net baryon density is so small that its
influence on the pressure and thus on the developing flow pattern is
hardly noticeable. For simpli\-ci\-ty we can thus also parametrize the 
initial net baryon density $n$ as being proportional to $n_{\rm WN}$:
 \bea{nw}
   n(\bs;\tau_0;\bb) = K_n(\tau_0) \, n_{\rm WN}(\bs;\bb)\,.
 \eea
Equations (\ref{sw}) and (\ref{nw}) together define the first model
for the initialization, sWN, to be studied below.

The implementation of the initialization (\ref{sw}) into the hydrodynamic
computer code is complicated by the fact that in our formulation the 
hydrodynamic equations propagate the fields $n(x)$ and $e(x)$ (and not 
$s(x)$). The initialization (\ref{sw}) thus requires the additional step 
of computing $e(\bs;\tau_0;\bb)$ from $s(\bs;\tau_0;\bb)$ by solving an 
implicit equation resulting from the thermodynamic identity 
 \bea{thid}
   T(e,n)\, s = e + p(e,n) - \mu(e,n)\, n\,,
 \eea
where $T(e,n)$, $p(e,n)$, and $\mu(e,n)$ are tabulated values for the 
temperature, pressure and baryon chemical potential corresponding 
to the selected equation of state. 


\subsubsection{Model \ {\rm eWN}}
\label{sec2a2}


In a hydrodynamic approach with fixed freeze-out conditions, it is
strictly speaking not possible that both the transverse energy 
$d\ET/dy$ and the charged multiplicity $d\Nch/dy$ scale 
linearly with $N_{\rm WN}$. During the hydrodynamic evolution thermal 
energy is converted into longitudinal and transverse flow energy, and 
the time available for this conversion increases with the number of 
wounded nucleons. Since the transverse flow increases the average 
transverse energy per particle, $d\ET/dy$ should rise more quickly 
with $N_{\rm WN}$ than $d\Nch/dy$. There is some indication for 
this to happen in the WA98 data from Pb+Pb collisions at the SPS 
\cite{WA98}, where $\ET/\Nch$ at midrapidity rises slightly 
from very peripheral to semiperipheral collisions and then saturates 
from semiperipheral to central collisions. However, the experimental
effect is small and not clearly statistically significant. 

Given this unclear experimental situation, and because it simplifies 
the initialization process, we assumed in our previous work 
\cite{KSH99,KSH00,KH3,HKHRV} that it is the initial {\em energy} density 
(and not the entropy density) which scales with the density of wounded 
nucleons in the transverse plane:
 \bea{ew}
   e(\bs;\tau_0;\bb) = K_e(\tau_0) \, n_{\rm WN}(\bs;\bb)\,.
 \eea
Equation (\ref{ew}) together with Eq.~(\ref{nw}) defines our second 
initialization model, eWN. The effect of flow on $\ET/\Nch$
at RHIC and its dependence on the initialization of the hydrodynamic 
evolution will be discussed in Section~\ref{sec3}. 

\subsection{Hard particle production}
\label{sec2b}


At higher and higher collision energies, one expects that hard 
collisions among quarks and gluons from the nuclear structure functions
become more and more important and eventually dominate secondary particle
production \cite{KLL87}. In this limit secondary particle production
is a result of incoherent parton-parton collisions, and each 
nucleon-nucleon collision contributes equally to the cross section.
Particle and energy production should then be related to the distribution
of the number of binary nucleon-nucleon collisions in the transverse
plane. It is given in terms of the nuclear thickness functions (\ref{TA})
by
 \bea{TAB}
   {dN_{AB}^{\rm coll}(\bs;\bb)\over d^2s} = \sigma\,
   T_A\left(\bs{+}\half\bb\right) T_B\left(\bs{-}\half\bb\right)\,,
 \eea
where $\sigma$ is the nucleon-nucleon cross section. The integral 
of (\ref{TAB}) is the nuclear overlap function, $T_{AB}(\bb) = 
\int d^2s\, T_A\left(\bs{+}\half\bb\right) T_B\left(\bs{-}\half\bb\right)$, 
which is normalized to $AB$: $\int d^2b\, T_{AB}(\bb){\,=\,}AB$. It gives 
the total number of binary collisions as a function of impact parameter, 
$\Nc(b)$, which is also shown in Fig.~\ref{F1}. 
 

\subsubsection{Model {\rm sBC}}
\label{sec2b1}


If the system of secondary particles thermalizes quickly via elastic 
collisions, their number density defines, up to a proportionality 
constant, the initial entropy density at the beginning of the 
hydrodynamic expansion:
 \bea{sc}
   s(\bs;\tau_0;\bb) = \tilde K_s(\tau_0) \,
   T_A\left(\bs{+}\half\bb\right) T_B\left(\bs{-}\half\bb\right)\,.
 \eea
Assuming that the initial net baryon density in the transverse 
plane can also be calculated perturbatively from the nuclear 
structure functions \cite{EK97}, we write
 \bea{nc}
   n(\bs;\tau_0;\bb) = \tilde K_n(\tau_0) \, 
   T_A\left(\bs{+}\half\bb\right) T_B\left(\bs{-}\half\bb\right)\,.
 \eea
As discussed in Section~\ref{sec2a1}, the results do not depend 
on whether we use (\ref{nc}) or (\ref{nw}). The combination of 
Eqs.~(\ref{sc}) and (\ref{nc}) defines model sBC for the initialization.


\subsubsection{Model {\rm eBC}}
\label{sec2b2}

Within the perturbative approach to particle production one can also 
argue that each nucleon-nucleon collision contributes with equal 
probability not only to the number of produced secondaries, but also to
the energy carried by them. This leads to the ansatz that the initial {\em
energy} (and not the entropy) density is proportional to the density of
binary collisions in the transverse plane: 
 \bea{ec}
   e(\bs;\tau_0;\bb) = \tilde K_e(\tau_0) \, 
   T_A\left(\bs{+}\half\bb\right) T_B\left(\bs{-}\half\bb\right)\,.
 \eea
This equation together with Eq.~(\ref{nc}) defines model eBC for the 
initialization. 

\subsection{The saturation model}
\label{sec2c}


In the saturation model \cite{EKRT00,EKT01} the production of partons 
becomes inhibited below a saturation scale $p_{\rm sat}$, determined
as the transverse momentum scale where the produced partons start to 
overlap transversally. The formation time of the QGP at each point 
$\bs$ is given by $\tau_{\rm sat}=1/p_{\rm sat}(\sqrt{s},A,B,\bs;\bb)$. 
The local initial energy density profile can then be worked out at the 
central slice as in \cite{ERRT01},
 \bea{e_sat}
 \nonumber
   e(\bs;\bb) &=& \frac{d\ET^{\rm pQCD}}{d^2s\,dz} \\
 \nonumber
   &=& T_A\left(\bs{+}\half\bb\right) T_B\left(\bs{-}\half\bb\right)\\
   &&\times 
   \sigma\langle \ET\rangle(\sqrt s, p_{\rm sat},\Delta y,A,B,\bs;\bb) 
   \cdot \frac{1}{\tau_{\rm sat}\Delta y},
 \eea
where $dz\approx \tau_{\rm sat}\Delta y$ within the central rapidity
unit $\Delta y$. The first $\ET$-moment of the minijet distribution,
$\sigma\langle\ET\rangle$ \cite{KLL87,EK97}, is computed in lowest
order pQCD as the first $\pt$-moment of the distributions of minijets
with $\pt\ge p_{\rm sat}$ and $y$ within $\Delta y$. The EKS98 shadowing 
\cite{EKS98} of the parton distributions is included and the NLO 
contributions to $\sigma\langle \ET\rangle$ \cite{ET00} are simulated 
by an approximate factor $K{\,=\,}2$, as also done in \cite{EKRT00,EKT01}.

In \cite{EKT01} a fully saturated system was considered by extending
the computation down to very low values of $p_{\rm sat}{\,=\,}0.5$ GeV, 
and neglecting the tails of the number (energy) density distributions at
large transverse distances. Now, for the hydrodynamic description, we
have to consider the tails also. To maintain the spirit of the local 
saturation model of \cite{EKT01}, we restrict the saturation to the 
regions where $p_{\rm sat}\ge 0.75$\,GeV and compute the tail of the 
energy density profile from Eq.~(\ref{e_sat}) with $p_{\rm sat}=0.75$\,GeV. 
Through this procedure, the multipli\-ci\-ty of \cite{EKT01} in central 
collisions is recovered (if computed as in \cite{EKT01}). For central 
collisions, the tail contributes only $7\%$ to the total multiplicity. 
We would like to emphasize that we have not tried to fit the saturation 
model to the centrality data but, rather, to keep the approach as close 
to the orginal local saturation idea \cite{EKT01} as possible.

As discussed in \cite{EKRT00}, from the energy per particle point
of view the system looks thermal already at saturation. The same can
be shown to hold also in the locally saturated system \cite{EKT01}.
Within the saturation framework it is therefore meaningful to switch on
the hydrodynamic evolution already at $\tau_{\rm sat}$ which now is a
local quantity.

\begin{figure}
  \begin{center}
    \epsfxsize 7cm \epsfysize 5.5cm \epsfbox{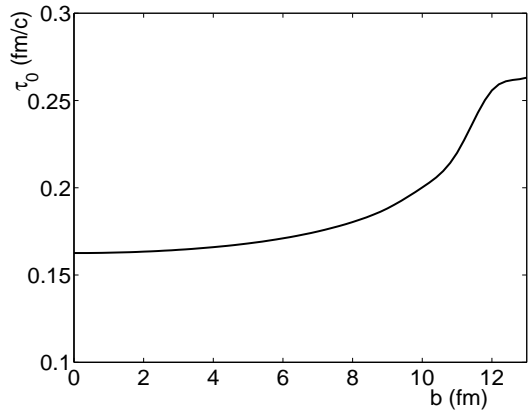}
  \end{center}
\caption{The formation and initial thermalization time $\tau_{\rm 
sat}(b){\,\equiv\,}\tau_{\rm sat}(\bs{\,=\,}0;\bb){\,=\,}1/p_{\rm sat}$ 
in the saturation model, as a function of impact parameter $b$.
}
\label{F2}
\end{figure}

For the hydrodynamic set-up used in this study, however, the density
profiles at a constant initial time are required. To circumvent this
problem, we consider two possibilities: first, for each impact 
parameter $\bb$, we simply use the earliest time, i.e. $\tau(\bs{\,=\,}0)$, 
as the initial time. These are shown in Fig.~\ref{F2}. Notice that 
the maximum initial time is $(0.75\,{\rm GeV})^{-1}{\,=\,}0.26$\,fm/$c$, 
and that a constant initial time is a good approximation for the central 
region and for small impact parameters. Alternatively, the energy 
densities computed from Eq.~(\ref{e_sat}) can be  evolved at each $\bs$ 
and at each $\tau_{\rm sat}(\bs;\bb)$ to $\tau_i{\,=\,}0.26$\,fm/$c$
assuming boost-invariant flow without transverse expansion:\footnote{We 
thank P.V. Ruuskanen for helpful discussions related to this point.}
 $e(\bs;\tau_i;\bb) = e\bigl(\bs;\tau_{\rm sat}(\bs;\bb);\bb\bigr)
\times\bigl(\tau_{\rm sat}(\bs;\bb)/\tau_i\bigr)^{4/3}$. We have checked, 
however, that the latter procedure leads only to a few percent increase 
in the central multipli\-ci\-ties relative to the former one. Therefore, we 
take in the following the initial time for the saturation model from 
Fig.~\ref{F2} and use $e(\bs;\tau_{\rm sat}(b);\bb)$ as the initial 
profile for the transversally expanding hydrodynamics, as discussed 
next.

\subsection{Initial energy density and spatial anisotropy}
\label{sec2d}

\begin{figure}
  \begin{center}
    \epsfxsize 7cm \epsfysize 5cm \epsfbox{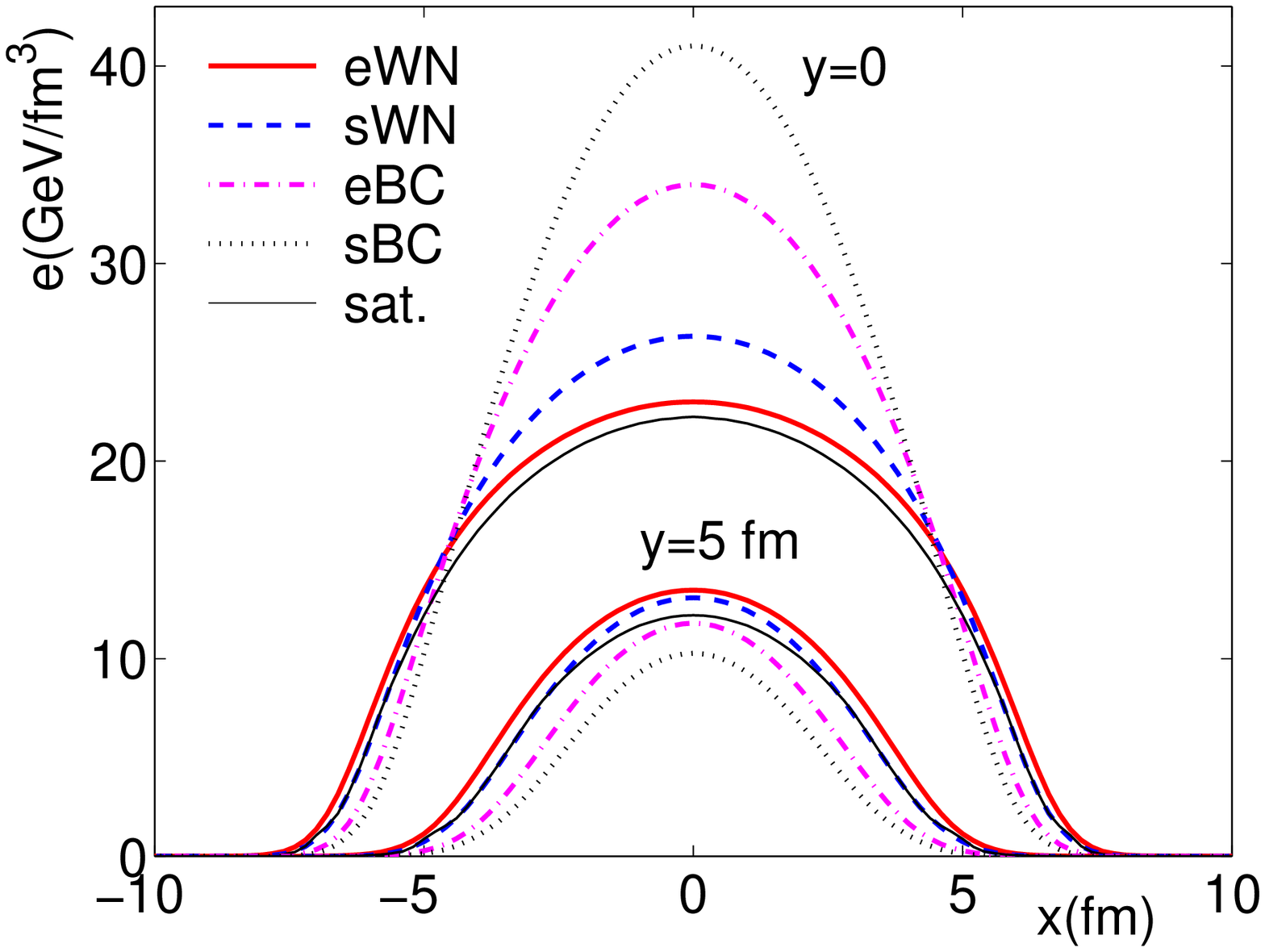}\\
    \epsfxsize 7cm \epsfysize 5cm \epsfbox{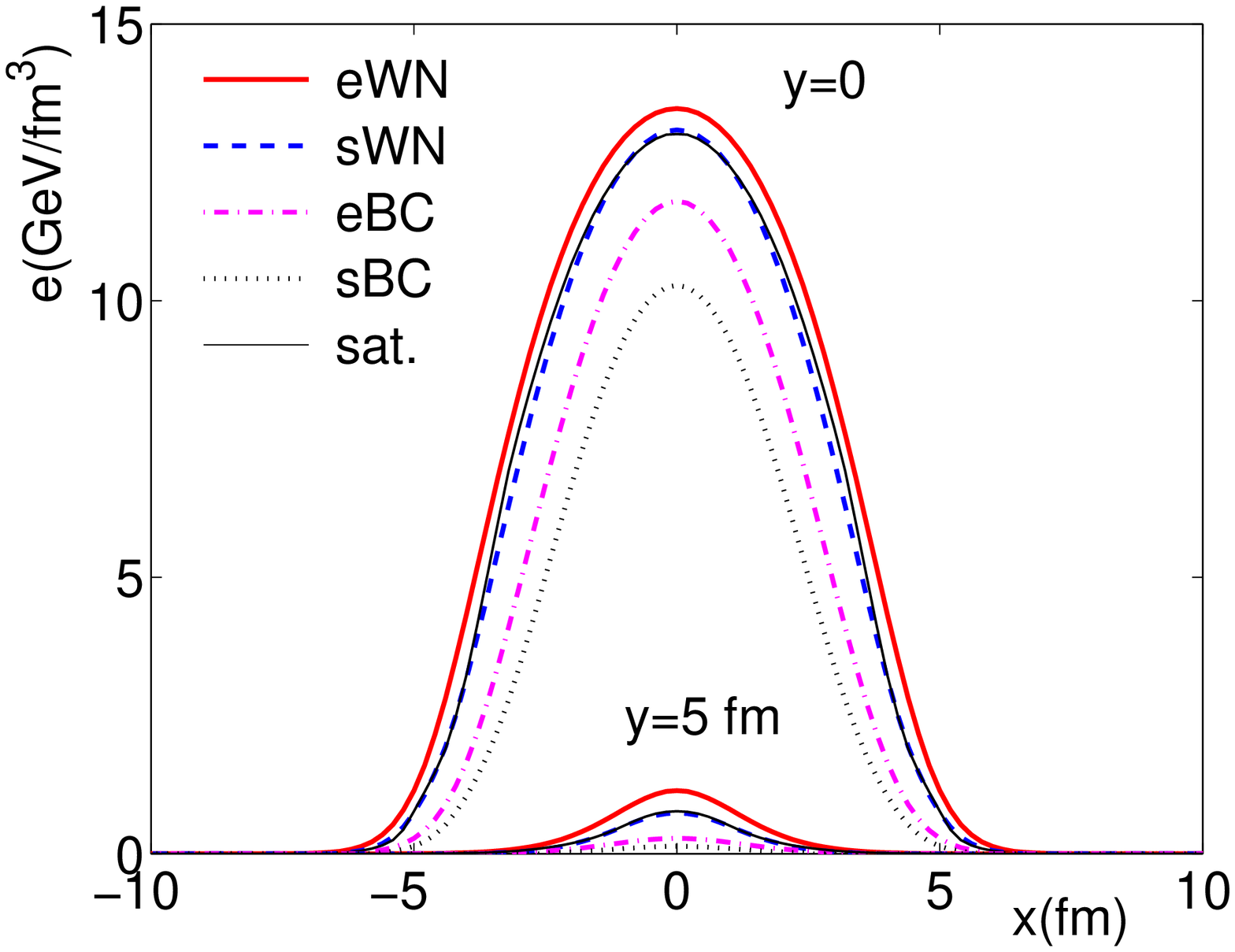}
  \end{center}
\caption{Initial energy density profiles in the transverse plane for 
the five different initialization models described in this Section. 
Top: $b{\,=\,}0$. Bottom: $b{\,=\,}10$\,fm. In each case two cuts in 
$x$-direction are shown, one for $y{\,=\,}0$ and one for $y{\,=\,}5$\,fm. 
For the saturation model (``sat.'') the profile was hydrodynamically 
propagated from the initial thermalization time assumed in that model
to $\tau_0=0.6$\,fm/$c$ where the hydrodynamic evolution was started 
for the other four models.   
}
\label{F3}
\end{figure}

Figure~\ref{F3} shows the initial energy density profiles for the 
different initialization models. ($x$ is the direction of the impact 
parameter $\bb$, while $y$ points perpendicular to the reaction 
plane.) For models sWN, eWN, sBC and eBC the profiles are normalized 
such that in each case the total charged multiplicity per unit 
pseudorapidity at $b{\,=\,}0$ is $d\Nch/dy{\,=\,}670$ at $y{\,=\,}0$. 
For model eWN \cite{KH3} this corresponds to 
$d\Nch/d\eta\vert_{\vert\eta\vert<1}{\,=\,}545$ for the $6\%$ most
central collisions (a bit less for the three other models). Within 
errors, this is consistent with the first published PHOBOS data 
\cite{PHOBOS}, but slightly below the more accurate recent data from 
PHENIX and PHOBOS \cite{PHENIX,Roland}. To preserve consistency with our 
previous publications \cite{KH3,HKHRV}, we decided against increasing 
our normalization of $d\Nch/d\eta$ to the new data since this would have 
implied retuning the initial conditions and freeze-out temperature in 
order to keep the spectra and elliptic flow unchanged. 

For the comparison with the other models in Fig.~\ref{F3}, the profile
obtained from the saturation model was evolved from the formation time
$\tau_{\rm sat}(b)$ (see Fig.~\ref{F2}) to $\tau_0{\,=\,}0.6$\,fm/$c$, 
the time at which the hydrodynamic evolution was started for the other 
four initializations. Within the line widths of Fig.~\ref{F3}, it did not
matter whether this scaling was done by assuming only boost-invariant
longitudinal expansion, $e(\bs,\tau_0){\,=\,}e(\bs,\tau_{\rm sat})\times 
 (\tau_{\rm sat}/\tau_0)^{4/3}$, or by including also the transverse 
expansion.

Since for a thermalized parton gas at nearly vanishing net baryon 
density $e{\,\sim\,}s^{4/3}$, Eq.~(\ref{sc}) gives distribu\-tions which are 
more sharply peaked at the origin than Eq.~(\ref{ec}). We will see that 
this results in a different centrality dependence of the total entropy 
per unit rapidity $dS/dy$ (which, for boost-invariant longitudinal 
expansion, is proportional to the integral of $s(\bs)$ over the 
transverse plane). Model eBC interpolates between model sWN, where 
$dS/dy{\,\sim\,}\Np$, and model sBC, where 
$dS/dy{\,\sim\,}\Nc{\,\sim\,}\Np^{4/3}$ (the latter proportionality 
was checked numerically to hold with excellent accuracy over the entire 
impact parameter range). This is similar to the HIJING model \cite{WG00} 
where the final charged particle rapidity density is a linear 
superposition of a soft component, scaling with $\Np$, and a hard 
component which scales with $\Nc{\,\sim\,}\Np^{4/3}$. In contrast to 
the mo\-dels here, however, HIJING has no rescattering among the produced 
particles and thus no collective flow. Finally, the saturation model is 
seen to be close to model eWN for central collisions, while for 
semiperipheral collisions it nearly coincides with model sWN.

\begin{figure}
  \begin{center}
    \epsfxsize 7cm \epsfysize 5cm \epsfbox{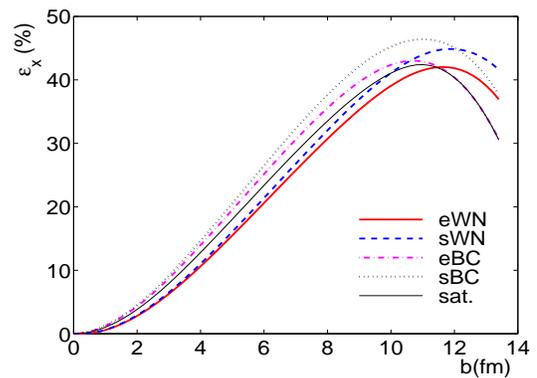}
  \end{center}
\caption{Initial spatial anisotropy as a function of impact parameter, 
for the different initializations.
}
\label{F4}
\end{figure}

Figure~\ref{F4} shows the initial spatial anisotropy
 \beq{eps_x}
   \epsilon_x(b) \equiv \delta(b) 
   = {\lda y^2{-}x^2 \rda \over \lda y^2{+}x^2 \rda}
 \eeq
as a function of impact parameter, evaluated with the initial transverse 
energy density as weight function, for the five initialization models 
studied here. At fixed impact parameter, $\epsilon_x$ varies by up to 
$20\%$. Since the hydrodynamic evolution maps the initial spatial 
anisotropy onto the final momentum-space, one expects from Fig.~\ref{F4}
that the impact parameter dependence of the elliptic flow $v_2$ should
show a similar sensitivity to the initialization. We will see
that, except for the saturation model, this model sensitivity is almost 
exactly cancelled by the corresponding variations in the impact parameter 
dependence of the produced charged particle multiplicity (i.e. entropy).


\section{Centrality dependence of multiplicity and transverse energy}
\label{sec3}


Using the initial transverse energy and baryon density profiles from
the previous section and the numerical code described in \cite{KSH00}, 
we solve the hydrodynamic equations for the transverse evolution of 
the reaction zone, assuming boost-invariant expansion in the 
longitudinal direction. We study only Au+Au collisions at 
$\sqrt{s}{\,=\,}130\,A$\,GeV. For the models eWN, sWN, eBC and sBC we 
use the same initial time $\tau_0{\,=\,}0.6$\,fm/$c$ as in \cite{KH3,HKHRV} 
to start the hydrodynamic expansion. For the saturation model the 
hydrodynamic expansion is started at $\tau_{\rm sat}(b)$ shown in 
Fig.~\ref{F2}. All calculations reported here are performed with 
EOS~Q, an equation of state with a first order phase transition 
from a hadron resonance gas to a non-interacting quark-gluon gas at 
$T_{\rm c}{\,=\,}164$\,MeV \cite{SHKRPV}, and freeze-out at 
$T_{\rm f}{\,=\,}120$\,MeV. The sensitivity of the spectra and 
elliptic flow to the EOS and freeze-out temperature were studied in 
\cite{KH3,HKHRV}. 


\subsection{Charged particle multiplicity}
\label{sec3a}

\begin{figure}
  \begin{center}
    \epsfxsize 7cm \epsfysize 5.3cm \epsfbox{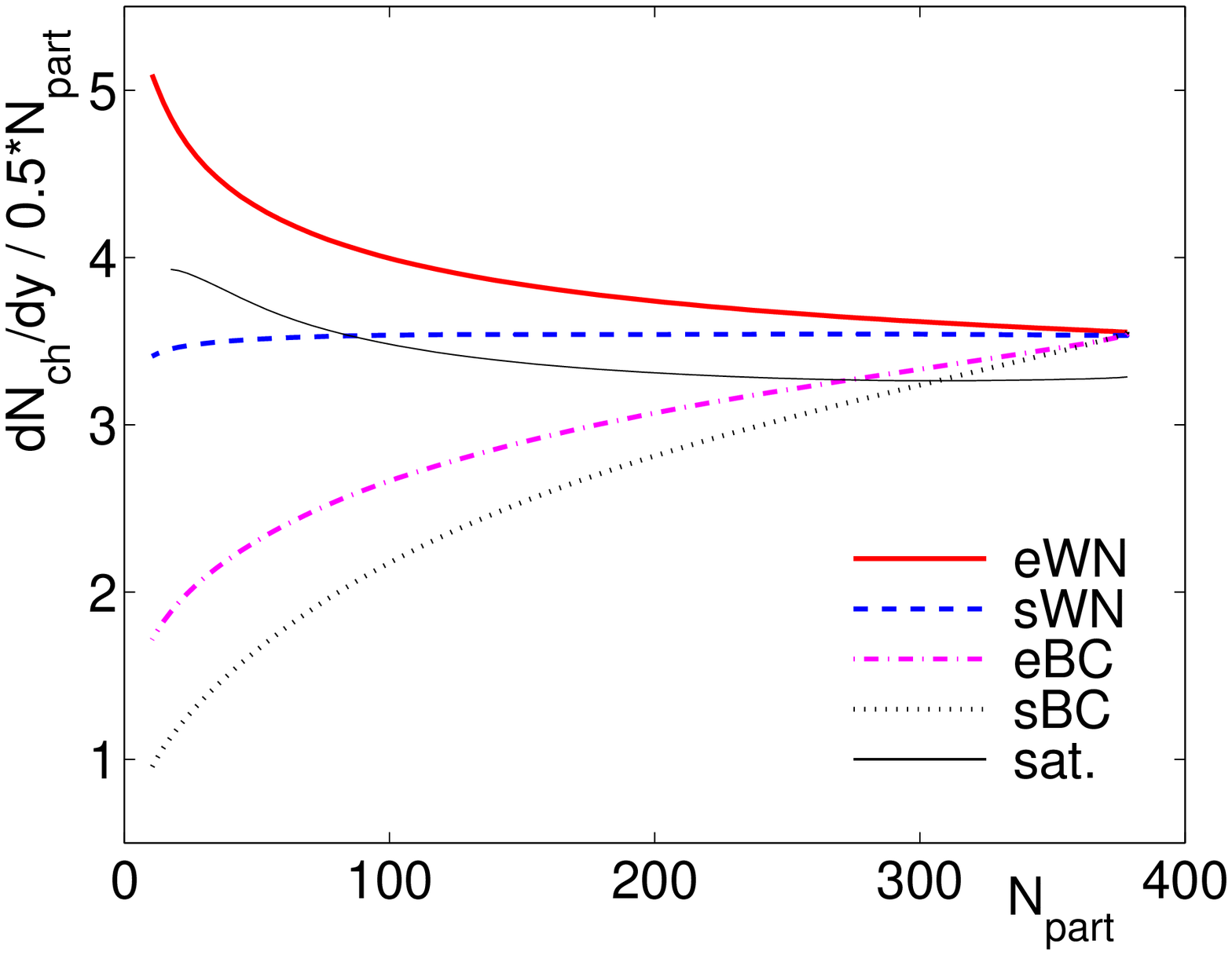}\\
    \epsfxsize 7cm \epsfysize 5.3cm \epsfbox{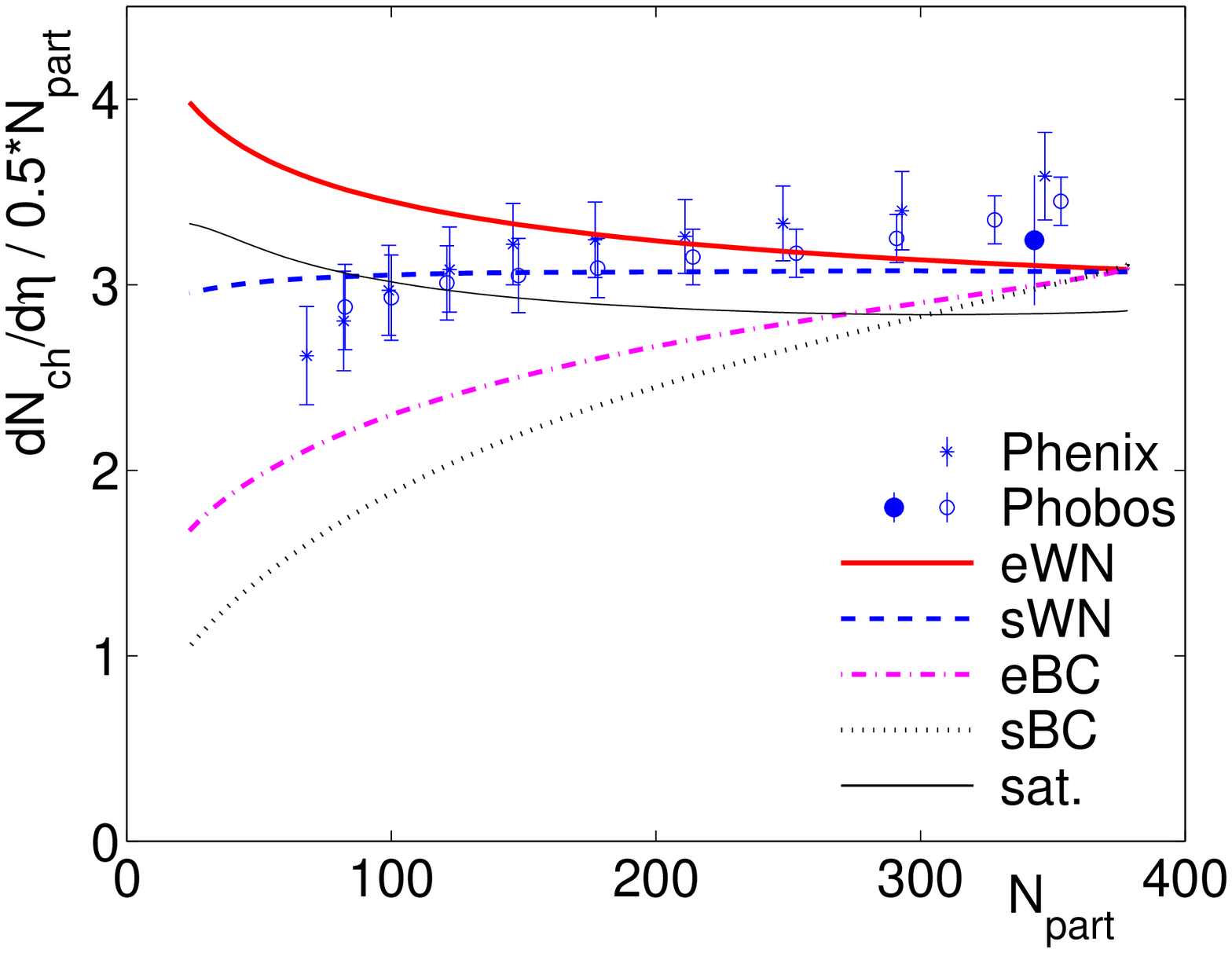}
  \end{center}
\caption{Charged particle yield per participating nucleon pair at 
midrapidity as a function of the number of participants. All curves 
were normalized to $d\Nch/dy=670$ for central ($b{\,=\,}0$)
collisions (see discussion below Fig.~\protect\ref{F3}). The top 
panel gives the rapidity, the bottom panel the pseudorapidity density. 
The data are taken from Refs.~\protect\cite{PHOBOS,PHENIX,Roland}.
}
\label{F5}
\end{figure}

Figure~\ref{F5} shows the final charged particle rapidity density per
participating nucleon pair resulting from the different initializations. 
In the bottom panel the rapi\-di\-ty density at midrapidity is converted 
to pseudorapidity density by the transformation
 \bea{conv}
   {d\Nch\over d\eta} = 
   \sum_{i\in{\rm charged}} \int \pt\, d\pt 
   {\pt\over \sqrt{m_i^2 + \bbox{p}^2}}\, 
   {dN_i\over dy\, \pt\, d\pt}\,,
 \eea
setting $y{\,=\,}\pL{\,=\,}0$. At fixed freeze-out temperature
$d\Nch/dy$ is a direct measure for the entropy density $dS/dy$, 
and the approximate constancy of the curve for model sWN thus 
reflects the approximate conservation of entropy by the hydrodynamic 
evolution. In fact, the slight increase of $(d\Nch/dy)/\Np$ for
small values of $\Np$ (Fig.~\ref{F5}, upper panel) can be traced back 
to a small amount of entropy production by deflagration shocks which 
arise during the hydrodynamic expansion stage as a consequence of the 
first order phase transition \cite{vH83} once the initial energy 
density in the center of the reaction zone increases above the 
critical value for QGP formation. 

Model eWN (solid line in Fig.~\ref{F5}) was used in Refs. 
\cite{KSH99,KSH00,KH3,HKHRV}, while model sWN was employed in 
\cite{O92,TLS00,TS99}. Fig.~\ref{F5} shows that both initialization 
models are disfavoured by the data, model eWN more so than mo\-del 
sWN. The saturation model produces a rather simi\-lar centrality dependence
of the charged multiplicity as model eWN. 

As expected, the saturation model results are close to those obtained
in Ref.~\cite{EKT01}. The careful reader notices, however, that the 
values $(d\Nch/dy)/(0.5\Np)$ for the saturation model shown in 
Fig.~\ref{F5} are ${\cal O}(10\%)$ lower than those shown in Fig.~4 
of \cite{EKT01}. This is due to a number of partially cancelling effects
\cite{ERRT01}: (i) Inclusion of the quark degrees of freedom increases
the initial entropy by about 20\% compared to \cite{EKT01}. (ii) In a 
realistic hadron resonance gas containing also heavy particles, at a 
freeze-out temperature of $T_{\rm f}{\,=\,}120$\,MeV each particle 
carries on average 4.87 units of entropy, instead of the 4 units assumed
in \cite{EKT01} (which is a good approximation in the ultrarelativistic
limit). This decreases the total multiplicity from that in \cite{EKT01}. 
(iii) Finally, the decay of unstable resonances changes the fraction of
charged particles from 2/3 \cite{EKT01} to $\approx 0.62$ of the total 
multiplicity. The net reduction effect for the final charged multiplicity 
rapidity density is about 10\% as mentioned.

The best representation of the {\em shape} of the data, even down to 
rather peripheral collisions (low $\Np$-values), seems to be given by 
model eBC, once appropriately renormalized to the new, slightly higher 
multiplicity densities in central collisions. Alternatively, one may try
a linear combination of the sWM and sBC parametrizations, with a large
sWN and a smaller sBC contribution \cite{WG00,KN00} (the latter is 
apparently absent at the SPS where the multiplicity scales linearly with
$\Np$). As discussed in Section~\ref{sec2c}, the main difference between 
the model eBC and the saturation model is that the $\bs$-dependence in 
eBC is contained only in the product $T_A\left(\bs{+}\half\bb\right) 
T_B\left(\bs{-}\half\bb\right)$ whereas in the saturation model 
\cite{EKT01} also the cross sections (and their moments) depend on 
$\bs$ through the local saturation scale $p_{\rm sat}(\bs)$. The data do 
not appear to support this particular implementation \cite{EKT01} of 
gluon saturation at the present RHIC energy of $\sqrt{s}{\,=\,}130\,A$\,GeV 
for the non-central collisions. In central collisions, however, parton 
saturation is not excluded as the dominant mechanism. We conclude that 
obviously the saturation region is now extended too far in transverse 
direction and that the eBC tails in the initial energy density profile 
(which were not included at all in \cite{EKT01}) should be given more 
weight than they have now. On the other hand, the data strongly indicate 
that the initial energy deposition process does involve a component which 
scales with the number of binary nucleon-nucleon collisions. We refrain 
from a discussion whether or not this indeed proves the onset of ``hard'' 
perturbative physics at RHIC \cite{WG00}. 


\subsection{Transverse energy per particle}
\label{sec3b}


Hydrodynamic flow is a result of the conversion of a fraction of the
thermal energy into collective flow kinetic energy, through work done 
by the thermodynamic pressure. As a consequence, the thermal energy per 
particle decreases as a function of time. If the reaction zone were to 
undergo boost-invariant expansion only along the longitudinal direction, 
$dS/dy\sim d\Nch/dy$ would be constant and $d\ET/dy$ would be proportional 
to the thermal energy density. The reduced thermal energy per particle
would thus be directly reflected in the finally observed transverse 
energy per charged particle, $(d\ET/dy)/(d\Nch/dy)$ \cite{GM84}. Since 
in more central collisions the thermodynamic pressure does longitudinal 
hydrodynamic work for a longer time, the finally observed ratio 
$(d\ET/dy)/(d\Nch/dy)$ should increase much more slowly with $\Np$ than 
the initial energy per particle \cite{DG00} established during the 
energy deposition process. In reality, the system undergoes not only
longitudinal, but also transverse expansion. The transverse collective 
flow adds a kinetic contribution to $\ET$ which reduces the diluting 
effect on $\ET/\Nch$ from the longitudinal expansion.
 
\begin{figure}
  \begin{center}
    \epsfxsize 8cm \epsfbox{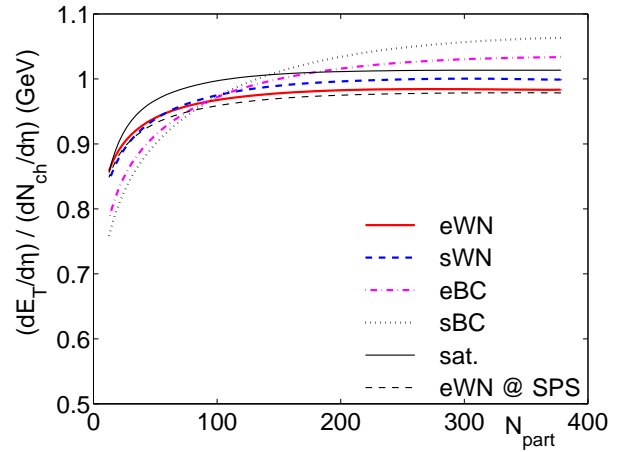}
  \end{center}
\caption{Average transverse energy per charged particle at midrapidity
as a function of the number of participants. Results are shown for all
five initialization models for Au+Au collisions at 
$\sqrt{s}{\,=\,}130\,A$\,GeV. In addition we plot the result from model 
eWN for Pb+Pb collisions at $\sqrt{s}{\,=\,}17\,A$\,GeV, using the 
initial conditions from Ref.~\protect\cite{KSH99,KSH00} (thin dashed line).}
\label{F6}
\end{figure}

This shows that the centrality dependence of $\ET/\Nch$ is an interesting 
observable which reflects the interplay of the centrality dependences 
of initial transverse energy and particle production, longitudinal work 
done by the hydrodynamic pressure, and radial flow. In Figure~\ref{F6}
we show the corresponding results from the hydrodynamic model for Au+Au
collisions at RHIC ($\sqrt{s}{\,=\,}130\,A$\,GeV) for the five different 
initialization models. For comparison we also show the centrality 
dependence for Pb+Pb collisions at the SPS, for model eWN with parameters 
adjusted to the measured spectra \cite{KSH99,KSH00}. Comparing the eWN 
curves for SPS and RHIC, one sees that the larger radial flow at RHIC has 
very little effect on the average $\ET$ per particle in the final state: the 
gain in transverse kinetic flow energy is largely compensated by more 
longitudinal work done at RHIC (smaller $\tau_0$, somewhat larger 
$\tau_{\rm f}$). The eWN curve at the SPS is consistent in shape with
the WA98 data shown in Fig.~14 of \cite{WA98}, but at the upper edge 
of their systematic error band.

At RHIC, the models eWN and sWN as well as the saturation model all
give similarly flat centrality dependences for the average $\ET$ per 
charged particle; they differ slightly in normalization and in how
steeply this ratio drops for very peripheral collisions. The data in 
Figure~\ref{F5} disfavour these initialization models. On the
other hand, Model eBC, which works better in Figure~\ref{F5}, exhibits 
a markedly stronger centrality dependence of $\ET/\Nch$, with a clearly 
visible slope below $\Np{\,\simeq\,}250$. (This is even more 
true for the initialization sBC, but for that model the corresponding 
curve in Fig.~\ref{F5} is too steep.) This reflects a similar strong 
centrality dependence of $(d\ET/dy)/(dS/dy)$ in the initial state, which 
is only partially offset by the increasing amount of longitudinal 
hydrodynamic work done as the collisions become more central. 

As recently pointed out in \cite{DG00}, a measurement of the ratio 
plotted in Fig.~\ref{F6} will be very helpful in complementing the 
information contained in Fig.~\ref{F5} and checking for the effects of 
longitudinal work. We note that non-equilibrium expansion scenarios like 
those studied in \cite{DG00}, where the matter performs less longitudinal 
work than predicted hydrodynamically, should lead to an even stronger 
dependence of $\ET/\Nch$ on $\Np$ than that shown in Fig.~\ref{F6}. In 
contrast, recent data from the PHENIX Collaboration \cite{PHENIX_ET}
exhibit the opposite tendency: they are flatter than the eBC curve in 
Fig.~\ref{F6} and, in fact, quantitatively consistent in magnitude and 
shape with the WA98 data at the SPS \cite{WA98}. It remains to be seen 
whether e.g. a linear combination of a large sWN with a small 
sBC component (which can be made to agree with the multiplicity data in
Fig.~\ref{F5} \cite{KN00}) leads to a sufficiently flat behaviour in 
Fig.~\ref{F6} to be consistent with the PHENIX data \cite{PHENIX_ET},
and whether the magnitude of $\ET/\Nch$ can be lowered by replacing the
unrealistic chemical composition from the hydrodynamical model by a more 
realistic one (our hydrodynamic model assumes that chemical equilibrium 
is preserved until kinetic freeze-out at $T_{\rm f}{\,=\,}120$\,MeV 
whereas preliminary STAR data indicate a much higher chemical freeze-out 
temperature of around 170--180 MeV \cite{QM01}).


\section{Radial and elliptic flow}
\label{sec4}

\vspace*{-0.2cm} 

\subsection{Radial flow}
\label{sec4a}

\vspace*{-0.2cm} 

\begin{figure}
  \vspace*{-0.9cm}
  \begin{center}
    \epsfxsize 8.65cm \epsfbox{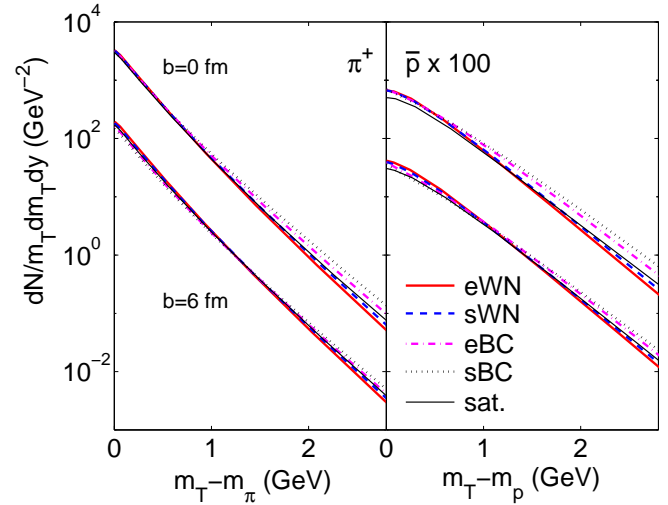}
  \end{center}
\caption
{Transverse mass spectra for positive pions and antiprotons, for 
central (upper set of curves) and semiperipheral collisions (lower 
set of curves, divided by 10 for clarity), for the five initialization 
models studied in this paper.
}
\label{F7}
\end{figure}

The single particle spectra for positive pions and antiprotons for 
the different initialization models are shown in Figure~\ref{F7}, for
two representative values of the impact parameter. 
For central ($b{\,=\,}0$) collisions, the systematics of the $\mt$-slopes 
shows that for models eWN, sWN, eBC and sBC the amount of radial flow 
created in the collision increases in the same order as the central initial 
energy density shown in Figure~\ref{F3}. The saturation model is an 
exception: it gives larger radial flow than both eWN and sWN although, 
at $\tau_0{\,=\,}0.6$\,fm/$c$, its central energy density is even below
that of model eWN. The reason is that for the saturation model the 
hydrodynamic evolution starts earlier and the build-up of transverse 
flow thus begins already at time $\tau_{\rm sat}(b)$ (see Fig.~\ref{F2}). 

The effect of the initialization on the single particle spectra is 
significant, and switching from model eWN used in \cite{HKHRV} to 
model eBC, which according to Figure~\ref{F5} provides a better fit 
to the centrality dependence of the charged multiplicity, considerably 
flattens the single particle spectra predicted in Ref.~\cite{HKHRV} 
for central collisions. This model sensitivity is weaker for 
semiperipheral collisions ($b{\,=\,}8$\,fm), again in agreement with 
Figure~\ref{F3} which shows that in this range the initial energy 
density profiles are rather similar for all the models. If 
hydrodynamics still applies to even more peripheral collisions, 
models eBC and sBC would there predict {\em steeper} spectra than 
models eWN and sWN.

\vspace*{-0.3cm} 

\subsection{Elliptic flow}
\label{sec4b}

\vspace*{-0.2cm} 

\begin{figure}
  \begin{center}
    \epsfxsize 7cm \epsfysize 5.5cm \epsfbox{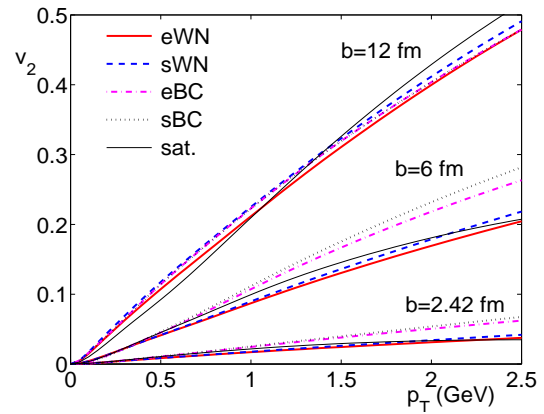}
  \end{center}
\caption{The $\pt$-differential elliptic flow $v_2(\pt)$ for charged
hadrons, at three representative impact parameters and for the five 
initialization models studied in this paper.}
\label{F8}
\end{figure}

Figure~\ref{F8} shows that for the four models eWN, sWN, eBC, and sBC 
the differential elliptic flow $v_2(\pt)$ follows a similar pattern as 
the radial flow: the model with the largest initial central energy 
density produces the strongest elliptic flow. The saturation model 
does not follow this systematics, which again is presumably due 
%
%
\noindent
to the earlier start of the hydrodynamic evolution in this model. The 
impact parameter dependence of this model sensitivity is now controlled 
by the initial spatial anisotropy $\epsilon_x$. According to Figure~\ref{F4},
for most of the impact parameter range it satisfies the hierarchy 
$\epsilon_x({\rm eWN}) < \epsilon_x({\rm sWN}) < \epsilon_x({\rm sat}) 
< \epsilon_x({\rm eBC}) < \epsilon_x({\rm sBC})$, and this is clearly 
reflected in the slopes of $v_2(\pt)$ for $\pt<2$\,GeV/$c$, as shown in 
Figure~\ref{F8}. Again, the effect is significant and can reach up to
$20\%$. At large impact parameters, $b>10$\,fm, and for $\pt>2$\,GeV/$c$ 
this simple ordering is broken.

\begin{figure}
  \begin{center}
    \epsfxsize 7cm \epsfbox{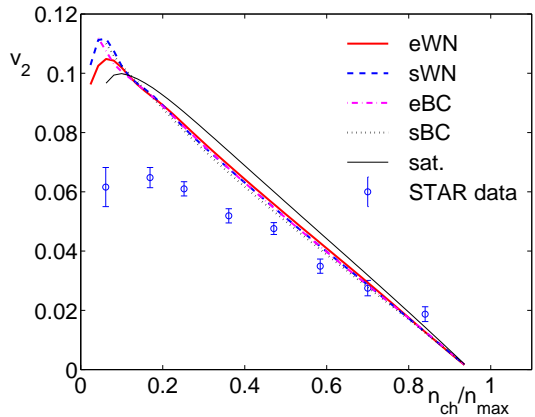}
  \end{center}
\caption{The $\pt$-integrated elliptic flow $v_2$ for charged hadrons
as a function of midrapidity charged multiplicity density, for 
different initialization models. The data are taken from 
Ref.~\protect\cite{STAR00}. (For a discussion of the horizontal 
axis $n_{\rm ch}/n_{\rm max}$ in theory and experiment, we refer to 
Refs.~\protect\cite{STAR00,KH3}.)
}
\label{F9}
\end{figure}

In view of these results, it is somewhat of a pleasant surprise to 
see in Figure~\ref{F9} that the $\pt$-integrated elliptic flow $v_2$
for all charged hadrons, when plotted as a function of the midrapidity
density of charged particle multiplicity as an experimentally easily
accessible centrality measure, shows very little sensitivity to the 
initialization model. The models eWN, sWN, eBC, and sBC yield almost 
identical results, and only for the saturation model $v_2$ is
slightly larger (except for the most peri\-phe\-ral collisions). Comparing 
the two extreme models eWN and sBC, one sees from Figure~\ref{F5} that, 
at fixed charge multiplicity, the latter favours smaller impact parameters 
which correspond to smaller $\epsilon_x$ and $v_2(\pt)$ (Figs.~\ref{F4} 
and \ref{F7}), but flatter single particle spectra which give more 
weight to larger $v_2$ at higher $\pt$ (Fig.~\ref{F6}). The net effect 
of this intricate interplay is an almost complete cancellation of these 
counteracting tendencies. 

\begin{figure}
  \begin{center}
    \epsfxsize 8.5cm \epsfysize 5.5cm \epsfbox{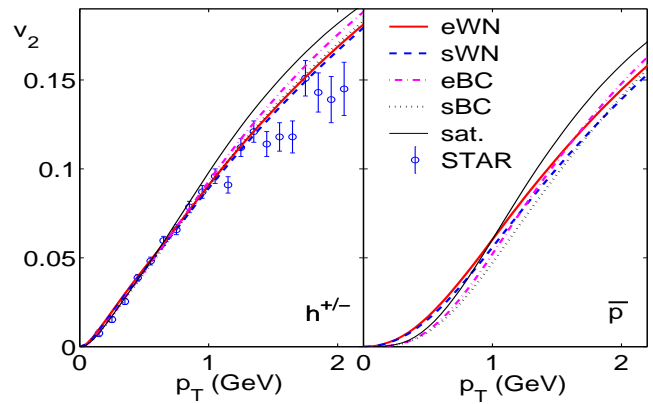}
  \end{center}
\vspace*{-0.2cm}
\caption{The differential elliptic flow $v_2(\pt)$ for minimum bias
events, for all charged hadrons (left) and for identified protons or
antiprotons (right). The curves correspond to different initialization 
models as indicated in the Figure. The data in the left panel are from 
Ref.~\protect\cite{STAR00}.
}
\label{F10}
\end{figure}

A similar cancellation happens for the differential anisotropy $v_2(\pt)$
for charged hadrons in minimum bias events (left panel of Figure~\ref{F10}):
 \bea{minbias}
   v_2(\pt) = 
   {\int b\,db\,v_2(\pt;b)\, {dN_{\rm ch}\over dy\,\pt\,d\pt}(b)
    \over  
    \int b\,db\,{dN_{\rm ch}\over dy\,\pt\,d\pt}(b)}\,.
 \eea
At fixed $b$, the models with smaller elliptic flow are weighted with
larger charged multiplicities, but slightly steeper spectra, and again
the net result is an almost miraculous cancellation of all sensitivities
to the initialization model for $\pt\,\lapp\,1.5$\,GeV/$c$. Above 
$\pt{\,=\,}1$\,GeV/$c$, the saturation model gives the largest $v_2(\pt)$.

This cancellation does not carry over to heavier particles. The right 
panel in Figure~\ref{F10} shows the differential elliptic flow $v_2(\pt)$ 
for identified protons and antiprotons, which is seen to exhibit 
more significant variations as the initialization of the hydrodynamic
model is changed. Generically, the model which gives the largest radial
flow at small impact parameters produces the smallest proton elliptic flow
at small values of $\pt$. This is qualitatively consistent with the
general analytic discussion of radial flow effects on the elliptic
flow for heavy particles presented in Ref.~\cite{HKHRV}. Again, the 
saturation model gives the largest elliptic flow of all studied 
initializations for protons with $\pt>1$\,GeV/$c$.

\section{Conclusions}
\label{sec5}


Within a hydrodynamic model with exact longitudinal boost invariance, we 
have shown that the centrality dependence of the production of particles, 
transverse energy and transverse flow is significantly influenced by the 
shape of the initial energy density profile in the transverse plane. This 
profile is intimately related to the nature of the primary particle
production process which converts beam energy into the matter forming the 
collision fireball. The available data for the charged multiplicity in 
Au+Au collisions at RHIC as a function of collision centrality are 
compatible with an initial energy deposition process involving a component 
which scales with the number of binary nucleon-nucleon collisions. They 
strongly disfavour the wounded nucleon parametrization used in earlier 
hydrodynamic simulations \cite{O92,KSH99,KSH00,KH3,HKHRV,TLS00,TS99}. In 
non-central collisions they are, at the present RHIC energy, also at 
variance with predictions from the gluon saturation model \cite{EKT01}. 
The best agreement with the multiplicity data at $\sqrt{s}{\,=\,}130\,A$\,GeV
\cite{PHOBOS,PHENIX,Roland} would be obtained by assuming that most of 
the initial thermalized {\em energy} density is proportional to the number 
of binary nucleon-nucleon collisions, with a smaller ``soft'' contribution 
proportional to the number of wounded nucleons. If one instead parametrizes 
the initial {\em entropy} density as a superposition of ``hard'' and 
``soft'' components, as implied by the approaches in \cite{WG00,KN00}, 
one needs a larger ``soft'' and smaller ($\sim 10\%$) ``hard'' component 
\cite{KN00}. A model with similar properties is obtainable also in the 
saturation approach considered here by limiting the saturation region to 
smaller transverse distances and thereby increasing the contribution from 
the tail which scales with the number of binary collisions. All these
models predict an average transverse energy per particle in the final 
state which rises slowly with increasing number of participating nucleons. 
A measurement of $\ET/\Nch$ as a function of $\Np$ is sensitive to the 
longitudinal work done during the hydrodynamic evolution and can be used 
to check the consistency of the hydrodynamic approach.

The centrality dependence of the $\pt$-averaged elliptic flow $v_2$
and the differential elliptic flow $v_2(\pt)$ for all charged particles 
from minimum bias collision events exhibit only minor sensitivity to 
the shape of the initial energy density profile. Thus, the conclusions 
extracted from earlier studies of elliptic flow using the wounded 
nucleon parametrization (which is now disfavoured by the new multiplicity 
data) remain valid. For identified heavy hadrons such as (anti)protons 
the differential elliptic flow $v_2(\pt)$ at low $\pt$ is smaller for 
the binary collision induced initial conditions than for the wounded 
nucleon parametrizations. This stems from the somewhat stronger radial 
flow created by the ``hard'' initializations, which should also be 
visible in the single-particle spectra.

In closing we note that the initialization models studied in the present
paper assume a one-to-one correspondence between impact parameter, number 
of participating nucleons, and initial energy density profile in the 
transverse plane. This neglects the possibility of strong event-by-event 
fluctuations in the initial density profile at fixed number of charged 
particles in the final state. As pointed out recently \cite{OAHK01},
strong event-by-event fluctuations in the initial energy density profile 
may result in sizeable event-by-event fluctuations of $v_2$ at fixed 
impact parameter. Our present study shows that small and smooth variations 
in the initial energy density profile at fixed impact parameter cause
effects on the particle multiplicity and on the radial and elliptic flow 
which compensate each other in such a way that $v_2(n_{\rm ch}/n_{\rm max})$ 
is almost unaffected. It would be interesting to study whether this 
compensation also survives the much larger fluctuations studied in 
Ref.~\cite{OAHK01}.

\bigskip\bigskip

\noindent{\bf Acknowledgements:} We thank P.V. Ruuskanen for several useful 
discussions. The work of P.K. was supported in part by the Deutsche 
Forschungsgemeinschaft. P.H. acknowledges financial support by the 
Director, Office of Science, Office of High Energy and Nuclear Physics, 
Division of Nuclear Physics, and by the Office of Basic Energy Sciences, 
Division of Nuclear Sciences, of the U.S. Department of Energy under Contract 
No. DE-AC03-76SF00098. K.J.E. and K.T. acknowledge financial support 
from the Academy of Finland. 



\end{document}